\documentclass[12pt]{article}

\usepackage{amsmath,amssymb,amsfonts,amsthm,dsfont,dcolumn}
\usepackage[margin=-5pt]{caption}
\usepackage{graphicx,wrapfig}
\usepackage{tensor}
\usepackage[hypertex]{hyperref}

\DeclareGraphicsExtensions{.pdf,.png,.jpg,.jpeg}

\newcolumntype{d}{D{.}{.}{0}}

\newlength{\textlength}
\newlength{\overlinelength}

\usepackage{color}

\newcounter{subequation}[equation]

\newcommand{\be}{\begin{equation}}
\newcommand{\ee}{\end{equation}}
\newcommand{\eel}[1]{\label{#1}\end{equation}}
\newcommand{\bea}{\begin{eqnarray}}
\newcommand{\eea}{\end{eqnarray}}
\newcommand{\eeal}[1]{\label{#1}\end{eqnarray}}

\makeatletter

\def\thesubequation{\theequation\@alph\c@subequation}
\def\@subeqnnum{{\rm (\thesubequation)}}
\def\slabel#1{\@bsphack\if@filesw {\let\thepage\relax
   \xdef\@gtempa{\write\@auxout{\string
      \newlabel{#1}{{\thesubequation}{\thepage}}}}}\@gtempa
   \if@nobreak \ifvmode\nobreak\fi\fi\fi\@esphack}
\def\subeqnarray{\stepcounter{equation}
\let\@currentlabel=\theequation\global\c@subequation\@ne
\global\@eqnswtrue \global\@eqcnt\z@\tabskip\@centering\let\\=\@subeqncr

$$\halign to \displaywidth\bgroup\@eqnsel\hskip\@centering
  $\displaystyle\tabskip\z@{##}$&\global\@eqcnt\@ne
  \hskip 2\arraycolsep \hfil${##}$\hfil
  &\global\@eqcnt\tw@ \hskip 2\arraycolsep
  $\displaystyle\tabskip\z@{##}$\hfil
   \tabskip\@centering&\llap{##}\tabskip\z@\cr}
\def\endsubeqnarray{\@@subeqncr\egroup
                     $$\global\@ignoretrue}
\def\@subeqncr{{\ifnum0=`}\fi\@ifstar{\global\@eqpen\@M
    \@ysubeqncr}{\global\@eqpen\interdisplaylinepenalty \@ysubeqncr}}
\def\@ysubeqncr{\@ifnextchar [{\@xsubeqncr}{\@xsubeqncr[\z@]}}
\def\@xsubeqncr[#1]{\ifnum0=`{\fi}\@@subeqncr
   \noalign{\penalty\@eqpen\vskip\jot\vskip #1\relax}}
\def\@@subeqncr{\let\@tempa\relax
    \ifcase\@eqcnt \def\@tempa{& & &}\or \def\@tempa{& &}
      \else \def\@tempa{&}\fi
     \@tempa \if@eqnsw\@subeqnnum\refstepcounter{subequation}\fi
     \global\@eqnswtrue\global\@eqcnt\z@\cr}
\let\@ssubeqncr=\@subeqncr
\@namedef{subeqnarray*}{\def\@subeqncr{\nonumber\@ssubeqncr}\subeqnarray}

\@namedef{endsubeqnarray*}{\global\advance\c@equation\m@ne
                           \nonumber\endsubeqnarray}

\makeatletter \@addtoreset{equation}{section} \makeatother
\renewcommand{\theequation}{\thesection.\arabic{equation}}

\catcode`\@=11

\newcount\hour
\newcount\minute
\newtoks\amorpm \hour=\time\divide\hour by 60\minute
=\time{\multiply\hour by 60 \global\advance\minute by-\hour}
\edef\standardtime{{\ifnum\hour<12 \global\amorpm={am}
        \else\global\amorpm={pm}\advance\hour by-12 \fi
        \ifnum\hour=0 \hour=12 \fi
        \number\hour:\ifnum\minute<10
        0\fi\number\minute\the\amorpm}}
\edef\militarytime{\number\hour:\ifnum\minute<10 0\fi\number\minute}

\def\draftlabel#1{{\@bsphack\if@filesw {\let\thepage\relax
   \xdef\@gtempa{\write\@auxout{\string
      \newlabel{#1}{{\@currentlabel}{\thepage}}}}}\@gtempa
   \if@nobreak \ifvmode\nobreak\fi\fi\fi\@esphack}
        \gdef\@eqnlabel{#1}}
\def\@eqnlabel{}
\def\@vacuum{}
\def\marginnote#1{}
\def\draftmarginnote#1{\marginpar{\raggedright\scriptsize\tt#1}}
\def\draft{
        \pagestyle{plain}
        \overfullrule=2pt
        \oddsidemargin -.5truein
        \def\@oddhead{\sl \phantom{\today\quad\militarytime} \hfil
        \smash{\Large\sl DRAFT} \hfil \today\quad\militarytime}
        \let\@evenhead\@oddhead
        \let\label=\draftlabel
        \let\marginnote=\draftmarginnote
        \def\ps@empty{\let\@mkboth\@gobbletwo
        \def\@oddfoot{\hfil \smash{\Large\sl DRAFT} \hfil}
        \let\@evenfoot\@oddhead}

\def\@eqnnum{(\theequation)\rlap{\kern\marginparsep\tt\@eqnlabel}
        \global\let\@eqnlabel\@vacuum}  }

\renewcommand{\theequation}{\thesection.\arabic{equation}}

\def\appendix#1{
  \addtocounter{section}{-3}
  \setcounter{equation}{0}
  \renewcommand{\thesection}{\Alph{section}}
  \section*{Appendix \thesection\protect\indent \parbox[t]{11.15cm}
  {#1} }
  \addcontentsline{toc}{section}{Appendix \thesection\ \ \ #1}
  }
\def\l{\lambda}

\def\be{\begin{equation}}
\def\ee{\end{equation}}

\def \l {\lambda}

\def\l{\lambda}

\date{}
\begin{document}

\begin{titlepage}

\hfill MCTP-10-17 \\

\begin{center}

{\Large \bf Chaos in the Gauge/Gravity Correspondence}

\vskip .7 cm

\vskip 1 cm

{\large   Leopoldo A. Pando Zayas$^1$, and C\'esar A. Terrero-Escalante$^2$}

\end{center}

\vskip .4cm \centerline{\it ${}^1$ Michigan Center for Theoretical
Physics}
\centerline{ \it Randall Laboratory of Physics, The University of
Michigan}
\centerline{\it Ann Arbor, MI 48109-1120}

\vskip .4cm \centerline{\it ${}^2$  Facultad de Ciencias}
\centerline{ \it Universidad de Colima}
\centerline{\it Bernal D\'{\i}az del Castillo 340, Col. Villas San Sebasti\'an,
Colima}
\centerline{\it Colima 28045, M\'exico}

\vskip .4cm
\centerline{ \it }
\centerline{\it  }

\vskip 1 cm

\vskip 1.5 cm

\begin{abstract}
We study the motion of a string in the background of the Schwarzschild black hole in $AdS_5$ by applying the standard arsenal of dynamical systems. Our description of the phase space includes: the power spectrum, the largest Lyapunov exponent, Poincare sections  and basins of attractions. We find convincing evidence that the motion is chaotic. We discuss the implications of some of the quantities associated with chaotic systems for aspects of the gauge/gravity correspondence. In particular, we suggest some potential relevance for the information loss paradox.
\end{abstract}

\end{titlepage}

\section{Introduction}
The AdS/CFT correspondence \cite{Maldacena:1997re,Witten:1998qj,Gubser:1998bc,Aharony:1999ti} has been mainly used as a window into the physics of strongly coupled  field theories but it also provides an approach to some gravity mysteries. For example, the question of black hole radiation can be, in principle, reformulated as a process in a unitary field theory.

The fact that the quantum numbers of certain operators  or states in field theory can be well described by the corresponding classical string is the idea at the heart of  Regge trajectories where the hadronic relationship $J\sim M^2$ is realized by a spinning string. This idea originated in the 1960's and certainly predates the AdS/CFT correspondence, it has found many concrete realizations in this new context. More generally, the AdS/CFT correspondence provides a dictionary that identifies states in string theory with operators in field theory. One of the most prominent examples is provided by the Berenstein-Maldacena-Nastase (BMN) operators. The BMN operators \cite{Berenstein:2002jq} can be described as a string moving at the speed of light in the large circle of $S^5$, the operator corresponding to the ground states is given by ${\cal O}^J=(1/\sqrt{JN^J}){\rm Tr}\,\,Z^{J}$. Another interesting class of operators which are nicely described as semiclassical strings in the $AdS_5\times S^5$ background are the Gubser-Klebanov-Polyakov (GKP) operators discussed in \cite{Gubser:2002tv}. They are natural generalizations of twist-two operators in QCD and in the context of ${\cal N}=4$ supersymmetric Yang-Mills they look like ${\rm Tr}\Phi^I\nabla_{(a_1} \ldots \nabla_{a_n)}\Phi^I$. A very important property of these operators is that their anomalous dimension can be computed using a simple classical calculation  and yields a prediction for the result at strong coupling $\Delta-S=(\sqrt{\lambda}/\pi)\ln S$.

These, and similar semiclassical states, have been the main window into the spectrum of string theory in $AdS_5\times S^5$ with Ramond-Ramond (RR) fluxes and have given very valuable information about anomalous dimension of operators with large quantum numbers. This is particularly crucial in the absence of methods that lead to the full spectrum of string theory in backgrounds with RR fluxes.

One of the most important results of XIX century mechanics was the identification of the role of the {\it full} phase space for characterizing a system, rather than the role of single trajectories. In view of the success of individual classical trajectories of strings in the gauge/gravity correspondence, it is only natural that we pose the question of the meaning of the full phase space. As soon as we pose this question we have to reckon with the fact that solving for the trajectories in phase space, that is, reducing them to quadratures is almost never possible. Integrable trajectories, in this sense, are a zero measure set. Even more, there are a number of reasons to expect the deterministic dynamics of strings in $AdS$ backgrounds to be {\it chaotic}, where by chaotic we mean {\it moving as if ruled by chance}. Starting from the 1898 paper by Hadamard \cite{Hadamard}, geodesic flows of point-like particles on manifolds of constant negative curvature have been found to satisfy the so-called Anosov (U) condition \cite{anosov,Arnold}. Roughly speaking, a dynamical system satisfies the Anosov (U) condition if near an arbitrary fixed trajectory the behavior of the neighboring trajectories with respect to the fixed one is similar to the behavior of the trajectories close to a saddle.
This particular kind of instability is typical of chaotic dynamics.
AdS is a space of constant negative curvature and it is not difficult to show that, as in the Hadamard example, the corresponding Jacobi equation \cite{Arnold,Hawking:1973uf} indicates that the deviation between neighboring geodesics of a point particle undergoes an exponential divergence. Even if AdS is not, rigorously speaking, compact, its causal structure is such that null and timelike geodesics going to infinity are reflected back, remaining effectively confined \cite{Hawking:1973uf}.
This way, the dynamics of point particles in AdS may be very complex.
Furthermore, in a interesting paper \cite{Frolov:1999pj}, Frolov and Larsen showed that, even for the Schwarzschild black hole in asymptotically flat space, if we consider the scattering of ring strings instead of point particles, the corresponding geodesic flow appears to be chaotic.
Not surprisingly, in this paper we find that the problem of a ring string orbiting around the Schwarzschild black hole in asymptotically $AdS_5$ space is highly sensitive to the initial conditions. This is a trademark signature of chaos. We support this finding by using a standard set of indicators which includes the power spectrum, the Poincar\'e sections, the largest Lyapunov exponent and the basins of attraction.

Therefore, if chaos is likely to be a common feature of the dynamics of semiclassical strings in $AdS$ backgrounds, it is relevant to wonder what it implies for the dual quantum field theory.
Our aim in this paper is to take the first steps into expanding the dictionary of the AdS/CFT correspondence into the realm of chaotic systems.  We propose some potential interpretation on the field theory side of the sensitivity to the initial conditions which could be typical of the chaotic motion of classical strings.
As an application, we briefly discuss how this interpretation could be potentially relevant for the resolution of the {\it Information Loss Paradox}.

The organization of the paper is as follows. In section \ref{setup} we discuss the set up including the embedding of the string in the Schwarzschild black hole spacetime. We also present the Hamiltonian description of the system. Section \ref{numerics}  is concerned with a discussion of the numerical methods we use and the standard analysis of the solution. Section \ref{chaos} contains a study of various signatures that generically indicate chaotic behavior. In section \ref{field} we present an argument for the possible identification of the operators being described by the ring string configuration discussed in this paper as generalizations of the GKP operators. We also make an attempt to interpreting a quantity typical of dynamical systems in terms of the dual field theory, namely, the largest Lyapunov exponent.
We summarize our ideas in the last section.

\section{Ring String in Schwarzschild-$AdS_5$}\label{setup}
\begin{figure}
\caption{Four dimensional sketch of the ring string dynamics. }
\label{fig:Dibujo}
\begin{center}
\includegraphics[width=.85\textwidth]{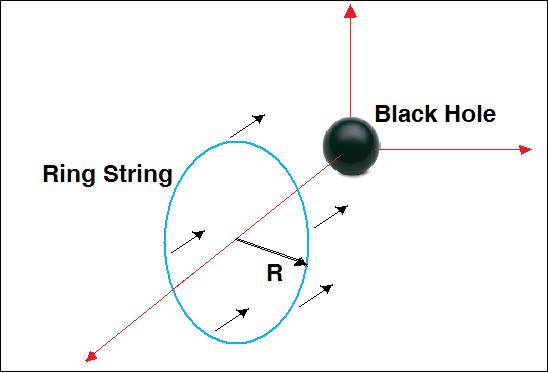}
\end{center}
\end{figure}

Looking for an example of chaos in AdS, we will describe the dynamics of a particular kind of string moving in five dimensions. Although the details will become clear later, in figure \ref{fig:Dibujo}
it is presented a simplified picture of the embedding of the string in spacetime.
We start with the Polyakov action
\be
{\cal L}=-\frac{1}{2\pi \alpha'} \sqrt{-g}g^{ab}G_{\mu\nu}\partial_a X^\mu \partial_b X^\nu,
\ee
where $G_{\mu\nu}$ is the spacetime metric of the fixed background, $X^\mu$ are the coordinates of the string, $g_{ab}$ is the worldsheet metric, the indices $a,b$ represent the coordinates on the worldsheet of the string which we denote as $(\tau, \sigma)$.
The background  metric $G_{\mu\nu}$ is of the form:
\be
\label{eq:metric}
ds^2 = -fdt^2 + \frac{dr^2}{f(r)}+ r^2 \left(d\theta^2 +\sin^2\theta d\psi^2 +\cos^2\theta d\phi^2\right), \qquad f(r)=1+\frac{r^2}{b^2}-\frac{w_4M}{r^2},
\ee
corresponding to the Schwarzschild black hole in $AdS_5$.
In order to explore the configuration sketched in figure  \ref{fig:Dibujo}, we consider a circular string whose world sheet is parametrized by $(\tau, \sigma)$ and is described by the following embedding
\be
\label{eq:embedding}
t=t(\tau), \qquad r=r(\tau),  \qquad \theta =\theta(\tau),  \qquad \phi=\phi(\tau), \qquad \psi =\alpha \sigma,
\ee
where $\alpha$ is a useful constant that will track the role of winding or more fundamentally the potential differences between strings and particles. In the conformal gauge, the Polyakov Lagrangian takes the form
\be
\label{Lagrangian}
{\cal L}=-\frac{1}{2\pi \alpha'} \bigg[f \dot{t}^2-\frac{\dot{r}^2}{f}-r^2\left(\dot{\theta}^2 +\cos^2\theta \dot{\phi}^2\right)+r^2\alpha^2 \sin^2\theta\bigg].
 \ee
It is worth noting that the term proportional to $\alpha$ enters with the same sign as the temporal direction, which suggests adding some energy to the system with respect to the particle limit.  The equations of motion are
\bea
\ddot{r}&=&-\frac{ff'}{2}\dot{t}^2 +\frac{f'}{2f}\dot{r}^2+rf\left(\dot{\theta}^2+\cos^2\theta\,\dot{\phi}^2-\alpha^2 \sin^2\theta\right), \quad  \dot{t}=E/f, \nonumber \\
\ddot{\theta}&=&-\frac{2}{r}\dot{r}\dot{\theta}-\sin\theta\cos\theta\left(\dot{\phi}^2 +\alpha^2\right), \quad r^2\cos^2\theta \, \dot{\phi}=l,
\eea
where $E$ and $l$ are constants of motion naturally related to the energy and angular momentum of this configuration, dot represents derivative with respect to $\tau$ and prime with respect to $r$. The nontrivial constraint from fixing the conformal gauge is
\be
G_{\mu\nu}\left(\partial_0 X^\mu\partial_0 X^\nu+\partial_1 X^\mu\partial_1 X^\nu\right)=0,
\ee
and in the above background takes the form
\be
\label{constraint}
-f \dot{t}^2 +\frac{1}{f}\dot{r}^2 +r^2 \left(\dot{\theta}^2+\cos^2\theta\,\,\dot{\phi}^2\right) +\alpha^2 \, r^2 \sin^2\theta=0.
\ee
The first three terms in the above expression can be thought of as describing the motion of a point particle with proper time $\tau$ moving on the space given by $(t,r,\theta,\phi)$ with a varying action given by the last term. A more suggestive way to write the constraint is
\be
\dot{r}^2 +r^2 f \dot{\theta}^2+\frac{f\, l^2}{r^2\cos^2\theta}+\alpha^2 \, r^2 f\sin^2\theta=E^2,
\ee
In which case it looks like the conservation of energy for a classical mechanical system.

Looking back at figure \ref{fig:Dibujo}, we note that in this configuration, the radius of the string changes in time as
\be
R(\tau)=r\sin\theta.
\ee
There are different modes of motion which can be characterized by the range of values of the radius $r$. The radius of the string, $R(\tau)$ above, also contributes to the intuition of the different modes. For example, on one possible mode of motion, the string can approach the black hole and get scattered, crossing over it. Then, it can get pulled back by the action of the black hole gravity, and so on. In this mode both quantities $r(\tau)$ and $\theta(\tau)$ oscillate in a way that $R(\tau)$ is always greater than the Schwarzschild black hole. Along these lines, and recalling the results in \cite{Frolov:1999pj}, we thus anticipate three modes of motion: i) the ring string remains forever crossing back and forward over the black hole ii) the ring string completes a number of these oscillations before collapsing into the black hole, ii) the ring string completes a number of these oscillations, before escaping to infinity.

\subsection{Hamiltonian Approach}

Let us consider the motion in the Hamiltonian formalism, this is particularly appropriate for the treatment of various quantities in the context of dynamical systems. We start with the Lagrangian (\ref{Lagrangian})
\be
{\cal L}=-\frac{1}{2\pi \alpha'}\bigg[f \dot{t}^2 - \frac{\dot{r}^2}{f} -r^2\left(\dot{\theta}^2 +\cos^2\theta \dot{\phi}^2\right) +r^2\alpha^2 \sin^2\theta\bigg].
\ee
The canonical momenta are
\bea
p_t&=& -\frac{1}{\pi \alpha'}f\dot{t}, \quad p_r= \frac{1}{\pi \alpha'}\frac{\dot{r}}{f}, \quad
p_{\theta}=\frac{1}{\pi \alpha'}r^2\dot{\theta}, \quad p_{\phi}= \frac{1}{\pi \alpha'}r^2\cos^2\theta\, \dot{\phi}, \nonumber \\
\eea
The Hamiltonian is of the form
\bea
H&=& \frac{\pi\alpha'}{2}\bigg[fp_r^2+\frac{p_{\theta}^2}{r^2}+\frac{p_{\phi}^2}{r^2\cos^2\theta}-\frac{p_t^2}{f}\bigg]
-\frac{1}{2\pi \alpha'}\,r^2\alpha^2 \sin^2\theta.
\eea
Let us consider the equations of motion following from our Hamiltonian. For the $t$ equation we have

\be
\dot{p}_t =0, \qquad \dot{t}=-\pi \alpha' \frac{p_t}{f}.
\ee
For the radial coordinate $r$
\bea
\dot{p}_r&=&-\frac{\pi\alpha'}{2}\frac{f'}{f^2}p_t^2 -\frac{\pi\alpha'}{2}f' p_r^2 +\pi\alpha'\frac{1}{r^3}p_{\theta}^2 +\pi\alpha'\frac{1}{r^3\, \cos^2\theta}p_{\phi}^2- \frac{1}{\pi\alpha'}\alpha^2 r \sin^2\theta, \nonumber \\
\dot{r}&=& \pi \alpha' f p_r.
\eea
The equation of motion for $\theta$ is
\bea
\dot{p}_\theta&=&-\pi\alpha'\frac{\sin\theta}{r^2\cos^3\theta}p_\phi^2-\phi\alpha'\alpha^2 r^2 \sin\theta \cos\theta, \nonumber \\
\dot{\theta}&=& \pi \alpha' \frac{p_\theta}{r^2}.
\eea
Finally, for the $\phi$ coordinate, we have
\be
 \dot{p_\phi}=0, \qquad \dot{\phi}=\pi \alpha' \frac{p_\phi}{r^2\cos^2\theta}.
\ee
Note that the constraint (\ref{constraint}) takes the simple form of
\be
\label{hamiltonianconstraint}
H=0.
\ee

The equations of motion for $t$ and $\phi$ can be eliminated from the system in favor of the corresponding constants of integration. Summarizing, our reduced Hamiltonian system is
\bea
\label{eq:sys}
H_r&=&\frac{\pi \alpha'}{2}\bigg[f\,p_r^2+\frac{1}{r^2}p_\theta^2 +\frac{l^2}{r^2\cos^2\theta}-\frac{E^2}{f}\bigg]+\frac{1}{2\pi \alpha'} \alpha^2\,r^2 \sin^2\theta, \nonumber \\
\dot{r}&=&\pi \alpha' \,\, f\,p_r, \nonumber \\
\dot{p_r}&=&-\frac{\pi\alpha'}{2}\frac{f'}{f^2}E^2-\frac{\pi\alpha'}{2}f'\,p_r^2 +\pi\alpha'\frac{1}{r^3}p_\theta^2 +\pi\alpha'\frac{l^2}{r^3\cos^2\theta}-\frac{1}{\pi\alpha'}\alpha^2 r \sin^2\theta, \nonumber \\
\dot{\theta}&=& \pi\alpha'\frac{1}{r^2}p_\theta, \nonumber \\
\dot{p}_\theta&=& -\pi\alpha'\frac{l^2}{r^2}\frac{\sin\theta}{\cos^3\theta}-\frac{1}{\pi\alpha'}\alpha^2 r^2 \sin\theta\cos\theta
\eea
subject to the constraint

\be
\label{eq:constraint}
H_r=0.
\ee
To develop our intuition further we can think of this constraint as we did from its Lagrangian analog (\ref{constraint}), in which case we simply have a point particle with a varying action. In other words, we could think of the above system as describing a point particle whose world line does not allow to be affinely reparametrized.

\subsection{Conserved quantities }

Classical conserved quantities of motion correspond to quantum numbers of the corresponding operator or state in the field theory. The worldsheet current of spacetime energy-momentum carried by the string is
\be
P^a{}_\mu=-\frac{1}{2\pi \alpha'} G_{\mu\nu} \partial^a X^\nu,
\ee
where $\alpha'$ is the string tension and the indices $a$ and $\mu$ were explained at the beginning of the section. Note that, interestingly
\be
P^\tau{}_t=\frac{E}{2\pi \alpha'}, \qquad P^\tau{}_\phi=\frac{l}{2\pi \alpha'}.
\ee
These quantities are related to the conformal dimension and spin of the corresponding operator. The quantity that is particular to our configuration is
\be
P^\sigma{}_\psi=-\frac{\alpha}{2\pi \alpha'}r^2 \sin^2\theta=-\frac{\alpha}{2\pi \alpha'}R^2,
\ee
where $R$ is the radius of the ring string. Thus, for this quantity we anticipate three possible asymptotic behaviors: i) oscillatory, ii) approaching zero or iii) approaching a constant finite value.

\section{Numerical treatment of the system}\label{numerics}
\begin{figure}
\caption{The reduced Hamiltonian as a function of proper time $\tau$ for different values of the error tolerance $\delta$ of the numerical solver. }
\label{fig:H}
\begin{center}
\includegraphics[width=1.\textwidth]{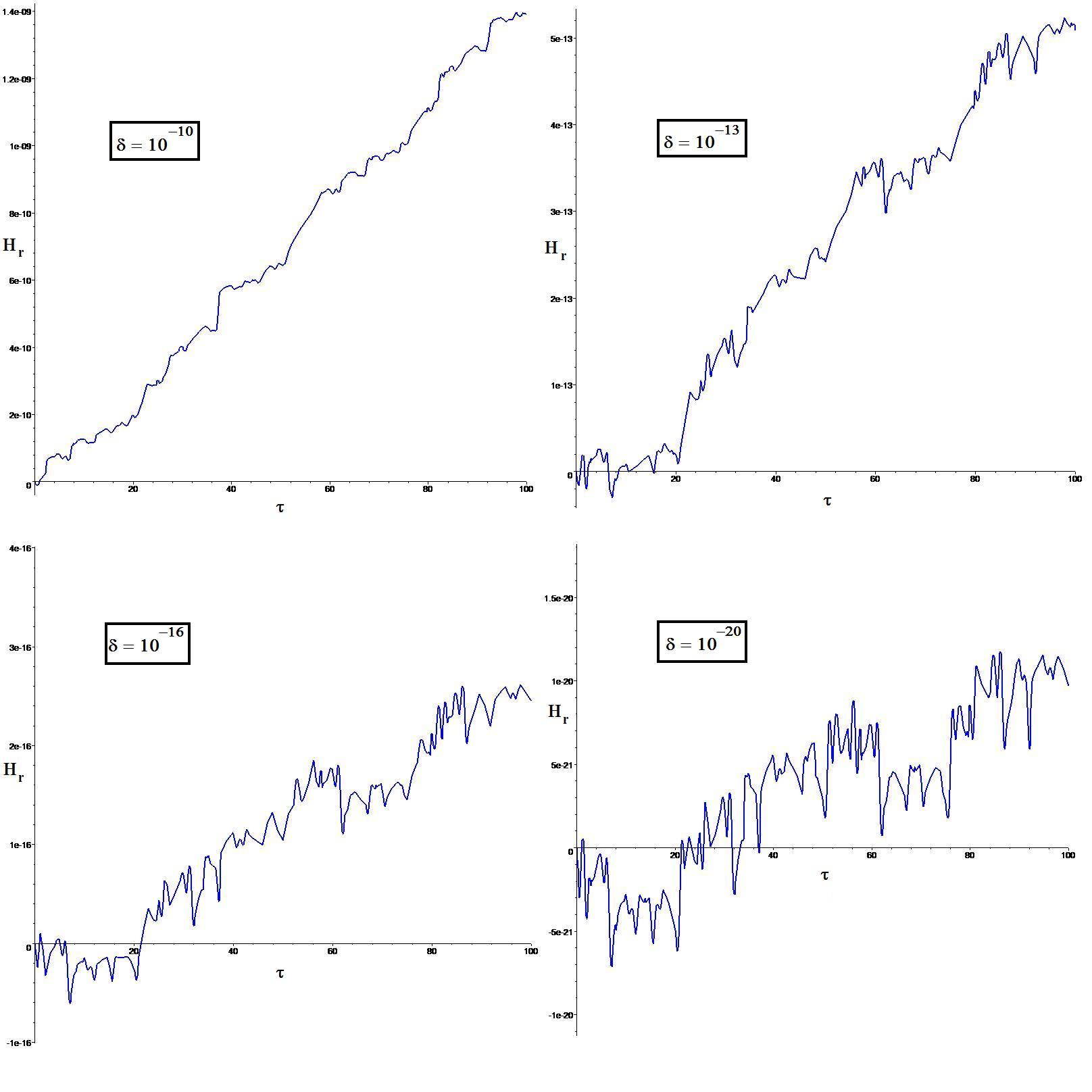}
\end{center}
\end{figure}

Let us briefly discuss aspects of the numerics that we feel are relevant for the reader. To solve the system (\ref{eq:sys}) we used a seventh-eight order continuous Runge-Kutta method which, thanks to its adaptive scheme, provides  great control upon the output accuracy. This accuracy can be assessed at any time by evaluating constraint (\ref{eq:constraint}). As it is shown in figure \ref{fig:H},
for a finite interval of time $\Delta \tau$ the numerical solver warrants that $|H_r|<\delta$, where $\delta$ is the (absolute and relative) error tolerance of the solver. The smaller the value of $\delta$, the larger the corresponding value of $\Delta \tau$. Since our aim is to detect chaos in this system, we will be typically considering the asymptotical behavior of the corresponding solutions which would imply the need of using small values of $\delta$.  Note that using constraint (\ref{eq:constraint}) to dynamically calibrate the trajectory is not recommended in this setup because we expect this system to be sensitive to the initial conditions and there will always be an error in the calibration coming from numerical truncation, that is, from solving $H_r=0$  to some numerical order. Fortunately, as we will show in section \ref{ssec:LLE} (see fig.\ref{fig:LLEE} and the corresponding table), our numerical results for very large $\tau$ are robust in the sense that they converge uniformly to given values as $\delta$ is decreased. This warrants that the first figures of our asymptotical numerical results obtained with reasonable values of  $\delta$ coincide with those to be obtained with smaller $\delta$. Our numerical experiments yield that to get six representative figures we need to fix  $\delta=10^{-12}$.

These numerical experiments and the results presented in the figures in this paper were obtained setting, in system (\ref{eq:sys}), $Mw_4=1$, $\alpha=1$, $b=10$, $\alpha^\prime=1/\pi$, $E=10$ and  $l=10$. These {\it ad hoc} values were chosen to yield a horizon radius near unity ($r_H=0.995085$) and, though a corresponding analysis must be done at some point,  we expect our results here to be generic with regards to the values of the parameters.

\begin{figure}
\caption{A solution $r(\tau)$ . }
\label{fig:rvst}
\begin{center}
\includegraphics[width=.85\textwidth]{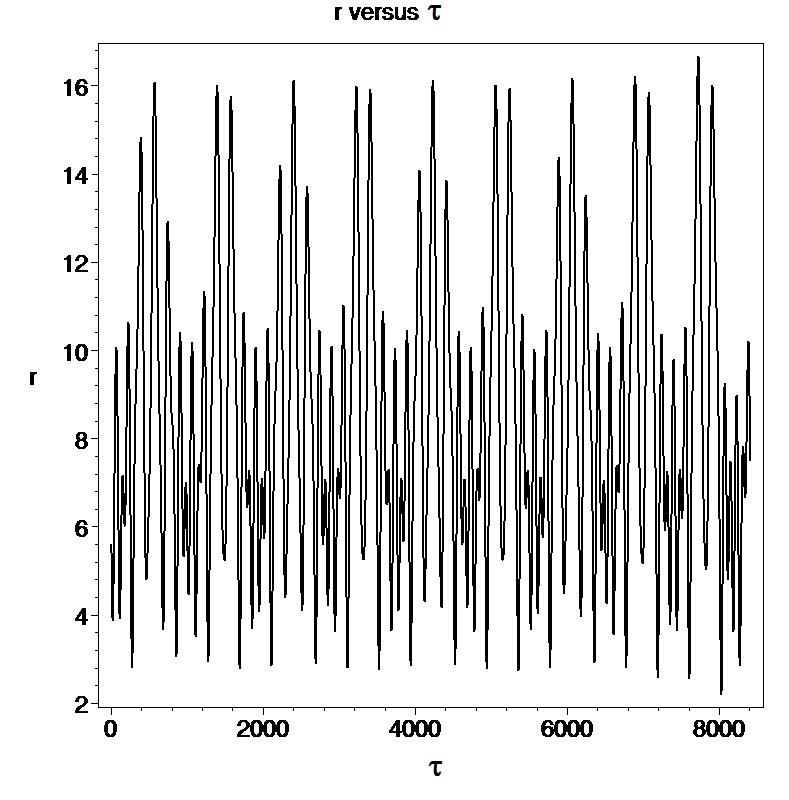}
\end{center}
\end{figure}

\begin{figure}
\caption{Phase curve corresponding to the solution plotted in \ref{fig:rvst}, projected into the ($r,p_r$) plane. }
\label{fig:prvsr}
\begin{center}
\includegraphics[width=.85\textwidth]{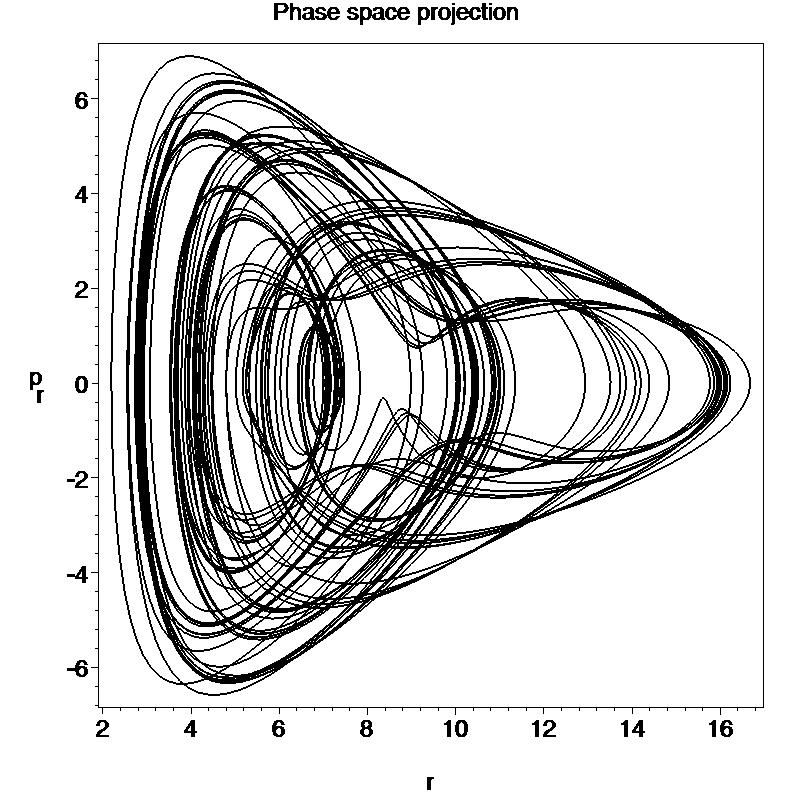}
\end{center}
\end{figure}

As our working example, in figure \ref{fig:rvst}
we present a plot of a solution for $r(\tau)$ obtained with the following set of initial conditions: $\{r(0)=5, p_r(0)=2.670855, \theta(0)=0, p_\theta=51.559927\}$. Though it looks regular, it is difficult to observe any definite pattern behind the amplitudes and frequencies of this oscillation. Moreover, in figure \ref{fig:prvsr}
we plotted the projection into the ($r,p_r$) plane of the phase space of the phase curve corresponding to this trajectory. It seems to fill densely a given region in the phase space. This complex behavior is typical of deterministic chaotic systems. The aim of the following section is to show that this is, indeed, the case for a ring string in the Schwarzschild black hole in $AdS_5$.

We end this section with some comments about our programs for the numerical computation of the dynamical indicators described below which were implemented in Maple following the recipes in references \cite{Enns,jcsprott, Ott}. They were checked and tuned using the canonical examples of a system of two decoupled harmonic oscillators and the well known H\'enon-Heiles system, which is a nonlinear nonintegrable Hamiltonian system, whose parameters allow both, for quasi-periodic as well as for chaotic behavior.

\section{Signatures of chaos}\label{chaos}

The na\"ive definition of deterministic chaos is that of a process that appears to proceed according to chance even though its behavior is in fact determined by precise laws (no randomness involved) \cite{Enns,jcsprott, Ott}.
In the case of interest to us, we will see that even if we have system (\ref{eq:sys}) which describes the geodesic motion of the ring string, the practical impossibility of exactly setting the initial conditions leads to the impossibility of predicting the final outcome for all the geodesics with initial conditions set inside a given ball in phase space, no matter how small the radius of this ball.

A rigorous mathematical definition of chaos is still lacking. In this direction there have not been fundamental advances since the works by Brown and Chua \cite{Brown:1996,Brown:1998}. In those papers they presented several examples leading to the conclusion that chaos is more a philosophical term than a mathematical one; no matter what definition is given for chaos, there is always some example of chaos which cannot be proven to be chaotic from the given definition.
Therefore, here we focus  on showing that our system has a number of properties that make it difficult to predict the large time behavior for any given geodesic when compared with the corresponding asymptotics of a neighboring geodesic at a given point of phase space. As was already mentioned, the interpretation of our results is done by comparison with the equivalent results obtained for a regular system of two decoupled harmonic oscillators and a quasi-periodic and a weakly chaotic configuration of the H\'enon-Heiles system.

\subsection{Invariant sets}

Since our emphasis is on the  large time behavior for any given geodesic, it is important to determine the possible (positive) invariant sets of the phase space of system (\ref{eq:sys}). For an invariant set with respect to the flow, every phase curve starting on it remains on it for ever. Particularly important examples are fixed points, limit cycles and Kolmogorov-Arnol'd-Moser (KAM) tori.

\subsubsection{Fixed points} \label{sssec:fp}
First we look for fixed points. Let us, for simplicity, focus on the values of parameters we are using in this paper. Note that, for finite $r$,  the right hand side of the equation for $r$  in system (\ref{eq:sys}) has real roots if and only if $p_r=0$. Similarly, the right hand side of the equation for $\theta$ is zero if and only if its corresponding conjugate momentum is zero. In turn, setting to zero the right hand side of the equation for $p_{\theta}$ give us $\sin\theta$ times a positive term, which leads to $\theta_*=n\pi$, where $n=0,1,2,\dots$. Using all these values in the equation for $p_r$ we obtain the algebraic equation $0.0099r_*^8-0.02r_*^6+0.02r_*^4+2r_*^2-1 = 0$ with only one real positive solution, $r_*=0.706009$. So this fixed point is located behind the event horizon. The eigenvalues of the Jacobian evaluated at this point are $(+80.598481 i, -80.598481 i, 14.190516,  -14.190516)$, indicating that this equilibrium is non-hyperbolic, possibly of saddle type, but clearly is not an attractor, because to be an attractor all the eigenvalues should have negative real parts.

Now, if we take the large $r$ limit of the right hand sides of system (\ref{eq:sys}), it can be determined that there is also a fixed point at $(r_*=\infty, p_{r*}=0, \theta_*=n\pi, p_{\theta *})$, for any finite $p_{\theta *}$. If $p_{\theta *}=0$, then for radial trajectories with $p_r>0$ ($p_r<0$), this fixed point is an attractor (repellor). If $p_{\theta *}\neq 0$, then at a large but finite value of $r$, the angle will eventually change, and as soon as this happen, it will result in a large angular momentum which can cause $p_r$ to change its sign from positive to negative and the geodesic to start a bounce toward the center. It can be verified that in this fixed point the Hamiltonian constraint is also satisfied. The conclusion seems to be that the set of trajectories escaping to infinity has zero measure.

\subsubsection{Limit cycles and KAM tori}

Our reduced system (\ref{eq:sys}) is Hamiltonian. If it were integrable, then it would have two first integrals in involution.
 Every level set of these two integrals is a 2-dimensional torus in the 4-dimensional phase space. This torus is invariant. Phase curves are densely wrapped around it;  generically both frequencies are functionally independent and they change from torus to torus. In general, not only both frequencies but also their ratio will change from torus to torus. Tori with only one frequency, i.e., limit cycles associated with asymptotically periodic motion, may also exist, but the probability of landing on such a torus by a random choice of an initial point in the phase space, is vanishingly small. The motion in the 4-dimensional phase space of an integrable system is generically quasi-periodic.
The KAM theorem \cite{kolmogorov,Arnold,Enns} states that for nearly integrable systems (integrable systems plus sufficiently small conservative hamiltonian perturbations), most invariant tori are not destroyed, but are only slightly deformed, so that in the phase space of the perturbed system, too, there are invariant tori densely filled with phase curves winding around them quasi-periodically, with a number of independent frequencies equal to the number of degrees of freedom. These invariant tori form a majority in the sense that the measure of the complement of their union is small when the perturbation is small. If one continues to perturb a KAM torus, it reaches a stage where the nearby phase space becomes self-similar (has fractal structure). At this point the torus is critical, and any increase in the perturbation destroys it. There are still quasiperiodic orbits that exist beyond this point, but instead of tori they cover cantor sets called {\it cantori}.  Thus, the transition to chaos in Hamiltonian systems can be thought of as the destruction of invariant tori, and the creation of cantori. For weakly chaotic systems it is common to find tori and cantori coexisting in the phase space \cite{Enns,jcsprott, Ott}.\\

\noindent {\bf Power spectrum}\\

\noindent
In the absence of randomness, the power spectrum of a signal from an integrable system allows us to determine if it is periodic or quasi-periodic and which are the relevant frequencies. In the corresponding spectrogram these frequencies are visible as vertical bars with heights proportional to the square of the amplitude related to each frequency in a Fourier decomposition of the signal. For instance, for the two decoupled harmonic oscillators the corresponding diagram consists of two vertical lines. Adding randomness implies adding new spurious vertical bars and modifying the power of the original ones. When (white) noise overpowers the (quasi)periodic signal, we are left with a flat spectrum over a defined frequency band. Signals from deterministic chaotic systems often are similar to those of noisy quasi-periodic systems, the power of the noise depending on how strong chaos is \cite{Enns,jcsprott, Ott}.

\begin{figure}
\caption{The power spectrum for the signal plotted in fig.\ref{fig:rvst}. }
\label{fig:Power}
\begin{center}
\includegraphics[width=.85\textwidth]{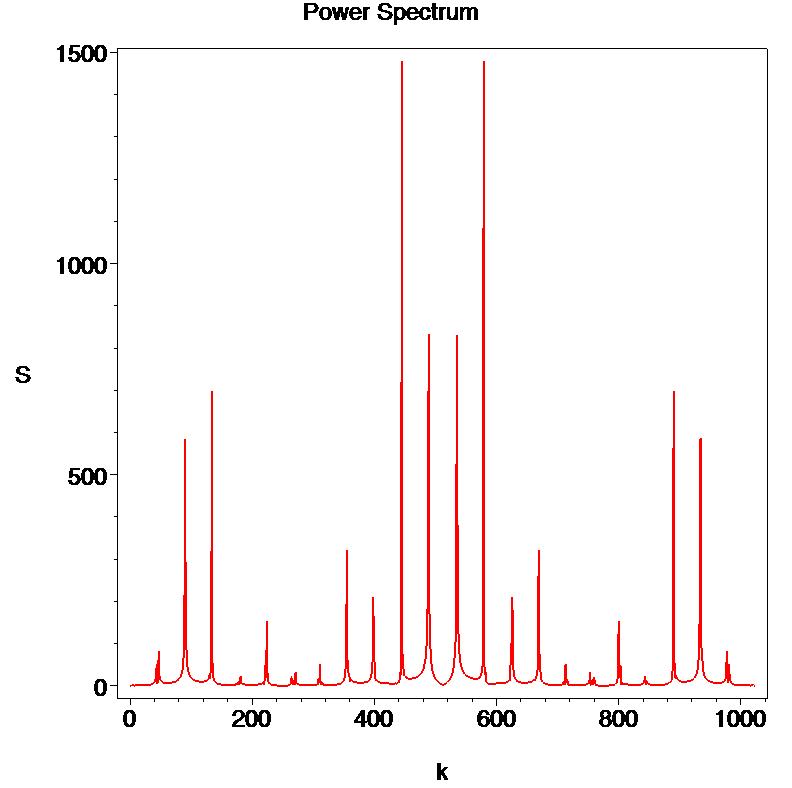}
\end{center}
\end{figure}

Figure \ref{fig:rvst} might give the false impression of periodicity. Indeed, it looks like two or three peaks appear at regular intervals of time $\tau$ which would correspond to a dominant frequency. However, in figure \ref{fig:Power}
 we present the (symmetric) spectrogram for the signal $r(\tau)$ plotted in figure \ref{fig:rvst}. As can be seen, there are around twelve distinct peaks in frequencies.
 Most of these peaks have slightly blurred basis which suggests that the system is weakly chaotic.\\

\noindent {\bf Poincar\'e sections}\\

\begin{figure}
\caption{The Poincar\'e section at $\theta=0$ for the signal plotted in fig.\ref{fig:rvst}. }
\label{fig:PS}
\begin{center}
\includegraphics[width=.85\textwidth]{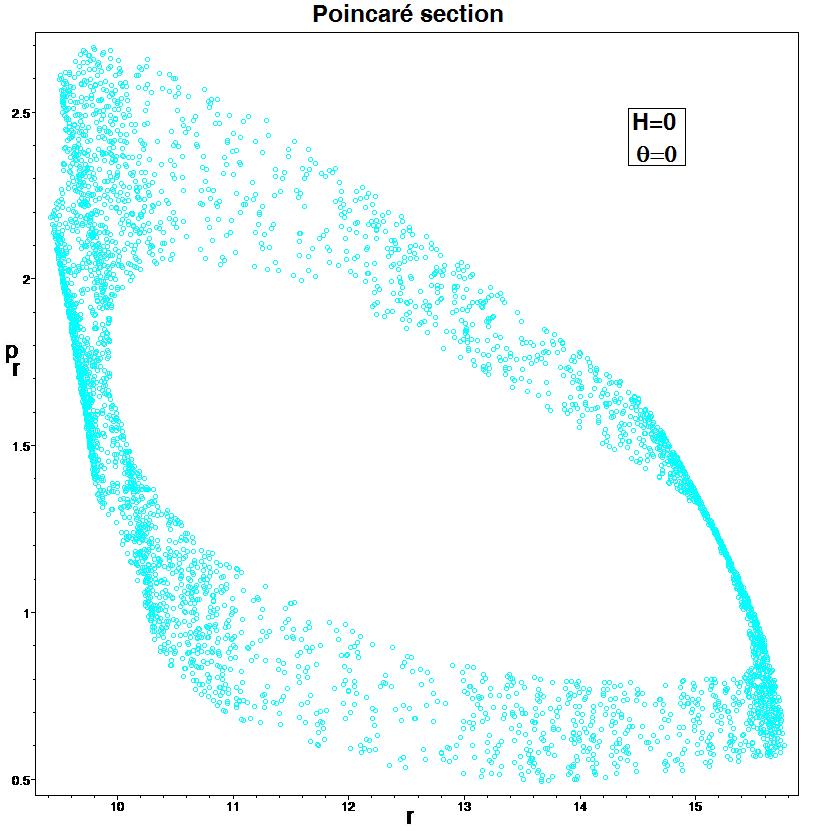}
\end{center}
\end{figure}

Since our phase space is four-dimensional, each energy level set is three-dimensional. We fix one such level set by setting $H_r=0$. This three-dimensional manifold, fibered by two-dimensional tori, can be represented in ordinary three-dimensional space as a family of concentric tori lying one inside another \cite{Arnold,Enns}.
Thus, if our system were integrable, for almost all initial conditions, a phase curve would densely fill an invariant 2-dimensional torus in the three-dimensional energy manifold.
It is then easy to construct a two-dimensional plane in the three-dimensional energy level set transversally intersecting the two-dimensional tori of our family (in a family of concentric circles in the model in three-dimensional Euclidean space). This plane is a co-dimension 2 Poincar\'e section. A phase curve beginning in such a plane returns to it after making a circuit around the torus. As a result we obtain a new point on the same circle (or any topologically equivalent curve) in which the torus intersects the plane. In this way there arises a mapping of the plane to itself, the Poincar\'e map. With this tool, the transition to chaos in Hamiltonian systems can be seen as the destruction of the circle in the Poincar\'e section, being substituted by a Cantor dust, the more diffuse, the stronger the chaos  \cite{Enns,jcsprott, Ott}.
In figure \ref{fig:PS}
we present a Poincar\'e section at $\theta=0$ for the solution plotted in figure \ref{fig:rvst}. As we can see, it confirms the result of the previous subsection of our system being weakly chaotic. Note also that we could take any point in this section as the initial conditions and we would obtain the same results presented in the figures of this paper.

\subsection{Largest Lyapunov exponent}\label{ssec:LLE}
One of the trademark signatures of chaos is {\it sensitive dependence on initial conditions}, which means that for any point $X\in \cal{M}$, there is (at least) one point arbitrarily close to $\cal{M}$ that diverges from $X$, where $\cal{M}$ is an invariant subset of the phase space and the divergence needs not to be exponential  \cite{Enns,jcsprott, Ott}.
Let us consider two points in phase space, $X_0$  and $X_0 + \Delta X_0$, each of which will generate an orbit in that space according to the flow given by (\ref{eq:sys}). These orbits are parametric functions of the proper time. If we use one of the orbits as reference orbit, then the separation between the two orbits will also be a function of time. In a neighborhood of  stable fixed points or attracting limit cycles, this separation decreases asymptotically with time, while it usually increases in the neighborhood of unstable fixed points or of repelling limit cycles. For orbits winding around KAM tori this distance eventually settles down to a constant value. Thus, this separation is also a function of the location of the initial value and has the form $\Delta X(X_0, \tau)$.  In cantori, the function $\Delta X(X_0, \tau)$ will behave erratically. It is thus useful to study the mean exponential rate of divergence of two initially close orbits using the formula
\be
\label{eq:lambda}
\lambda = \lim_{\tau \to +\infty} \lim_{ \Delta X_0 \to 0} \frac 1\tau \ln{\frac{\Delta X(X_0, \tau)}{ \Delta X_0}}.
\ee
This number, called the {\it characteristic Lyapunov exponent}, is the largest of a spectrum $\{\l_i\}$ which describes the evolution in time of a ball around an initial point in phase space.
The Lyapunov spectrum was rigorously defined by Osedelec \cite{Osedelec} in terms of the asymptotic behavior of the eigenvalues of the Jacobian of the vector field given by the right hand sides of the corresponding dynamical system.
For Hamiltonian systems, where the volume of phase space is preserved by the flow, $\sum{\l_i}=0$. What is important for us is the question of whether $\lambda$ is positive as it will indicate that two trajectories that start out very close in phase space diverge in time. If the motion takes place in a bound domain, a positive (largest) Lyapunov exponent is a good indication that the system may be chaotic.
Crucially for our study, it was already established that chaos, as characterized by a positive largest Lyapunov exponent, is coordinate invariant \cite{Motter:2003jm}. Moreover, recently this analysis was extended to include the class of transformations that do not preserve the boundedness of the orbits \cite{Motter:2009kv}. This may be relevant in the case of AdS.

\begin{figure}
\caption{The largest Lyapunov exponent as function of time for the signal plotted in fig.\ref{fig:rvst}. }
\label{fig:LLEvsT}
\begin{center}
\includegraphics[width=.85\textwidth]{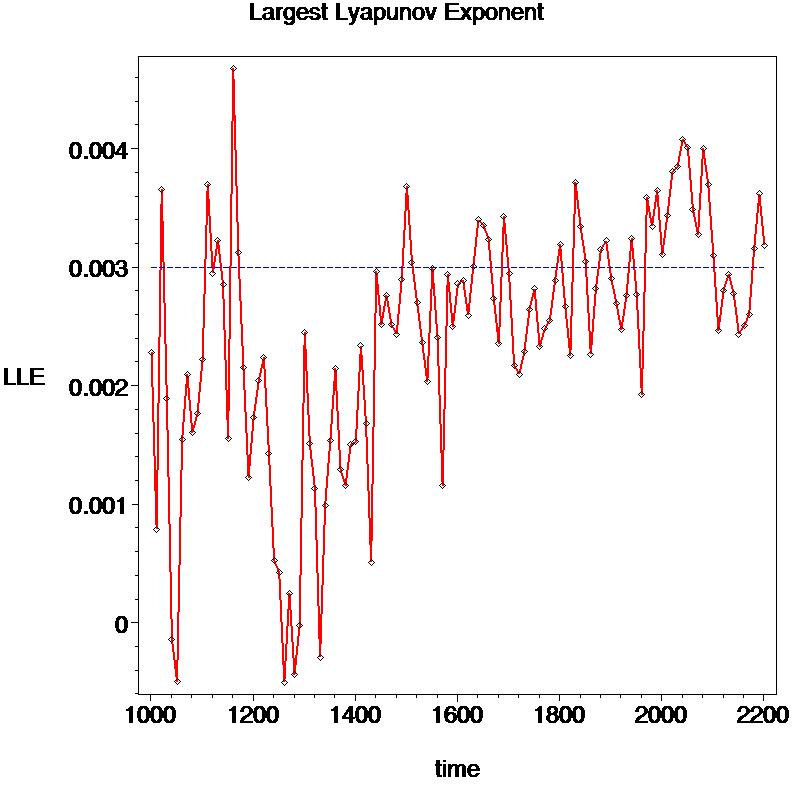}
\end{center}
\end{figure}

\begin{figure}
\caption{The largest Lyapunov exponent as function of $\delta$ for the signal plotted in fig.\ref{fig:rvst}. }
\label{fig:LLEE}
\begin{center}
\includegraphics[width=.85\textwidth]{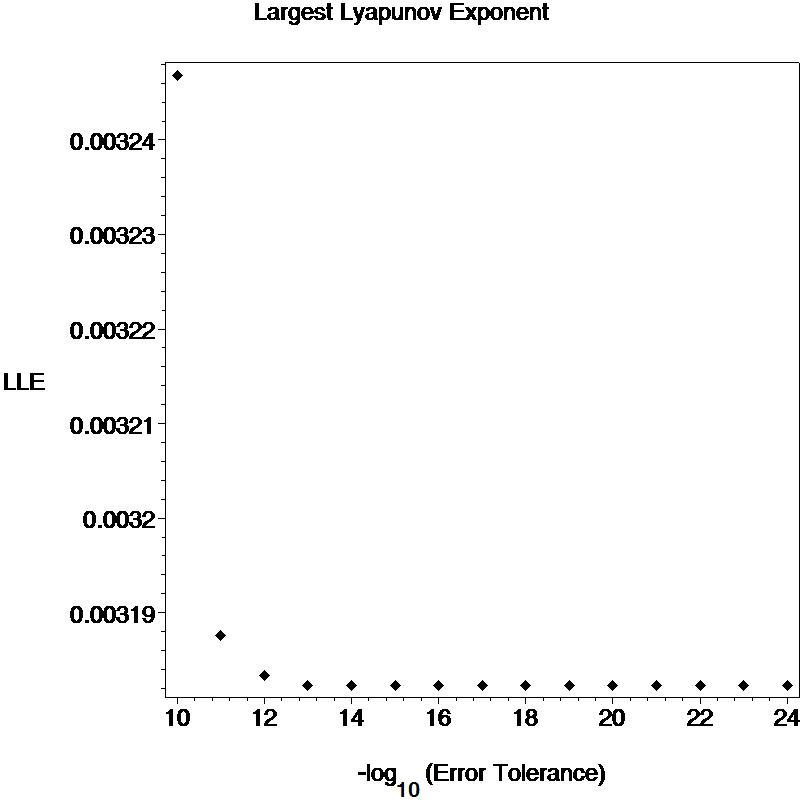}
\end{center}
\end{figure}

To calculate $\lambda$ we implemented the algorithm provided by Sprott \cite{jcsprott, sprott1}.
From the definition (\ref{eq:lambda}) we see that $\lambda$ is an asymptotic quantity.
To verify this numerically we should expect to observe that, as time $\tau$ is increased, $\lambda$ settles down to oscillate around a given value.
This is indeed observed in figure \ref{fig:LLEvsT}.
A comment about the precision of this calculation is in order.
As we have already mentioned, because of the cumulative numerical error, after some time the constrain $H_r=0$  is no longer satisfied within the error tolerance $\delta$.
The time interval where the constrain is satisfied is inversely proportional to $\delta$. So, we have repeated the calculation with increased precision.
The result is presented in the table and in figure \ref{fig:LLEE}.\\
{\large
\begin{tabular}{|c|c|}
	\hline
-$\ln$(Error Tolerance) & Largest Lyapunov Exponent \\
	\hline
10&\underline{0.003}2468399854926807     \\
11&\underline{0.003}1875419244818369     \\
12&\underline{0.00318}33206867751884     \\
13&\underline{0.00318}23399315688655     \\
14&\underline{0.00318233}81278734457     \\
15&\underline{0.00318233}75459681767     \\
16&\underline{0.003182337}4248104873     \\
17&\underline{0.0031823374}063400799     \\
18&\underline{0.00318233740}86545868    \\
19&\underline{0.003182337408}5976368    \\
20&\underline{0.0031823374085}858968   \\
21&\underline{0.003182337408585}4369   \\
22&\underline{0.0031823374085854}015    \\
23&\underline{0.003182337408585}3993\\
24&\underline{0.003182337408585399}1   \\
\hline
\end{tabular}}\\
We notice a numerical convergence in the results.
 This trend gives us confidence that the value we computed is truly indicative of a positive largest  Lyapunov exponent.
The small but positive value confirms that this is a case of weak chaos.

\subsection{Basins of attraction}
There is another signature of the unpredictability characteristic of chaotic systems which displays the sensitivity to the initial conditions. This signature is clearly defined in systems with more than one possible time-asymptotic behavior. It is possible, and even common, that at fixed values of parameters of a nonlinear system, more than one kind of outcome may be obtained. For instance, as we have already anticipated, in our case we have three possible outcomes: the geodesic crosses the event horizon (the ring string collapses into the black hole), the geodesic radially escapes to infinity (the ring string goes to infinity while its radius reaches a limit finite value) and the geodesics settles down to an oscillatory motion (the ring string crosses back and forth over the black hole forever, with its radius getting stretched and shrinked chaotically). The set of initial conditions (more precisely, the closure of this set) which eventually leads to each particular outcome is called its basin of attraction.
We may plot these basins by assigning a different color to each outcome, and then coloring all initial conditions according to their outcome. If the boundaries between these outcomes are smooth, then the dynamics is regular. Conversely, if the boundaries are fractal, the dynamics is chaotic \cite{McDonald:1985}.
A fractal is a non-differentiable structure and so cannot be removed by any differentiable coordinate transformation. Thus, relevantly for us, a fractal basin boundary also provides an observer independent signature of chaos \cite{Cornish:1997iw}.

\begin{figure}
\caption{Basins of attraction. }
\label{fig:BA0}
\begin{center}
\includegraphics[width=.85\textwidth]{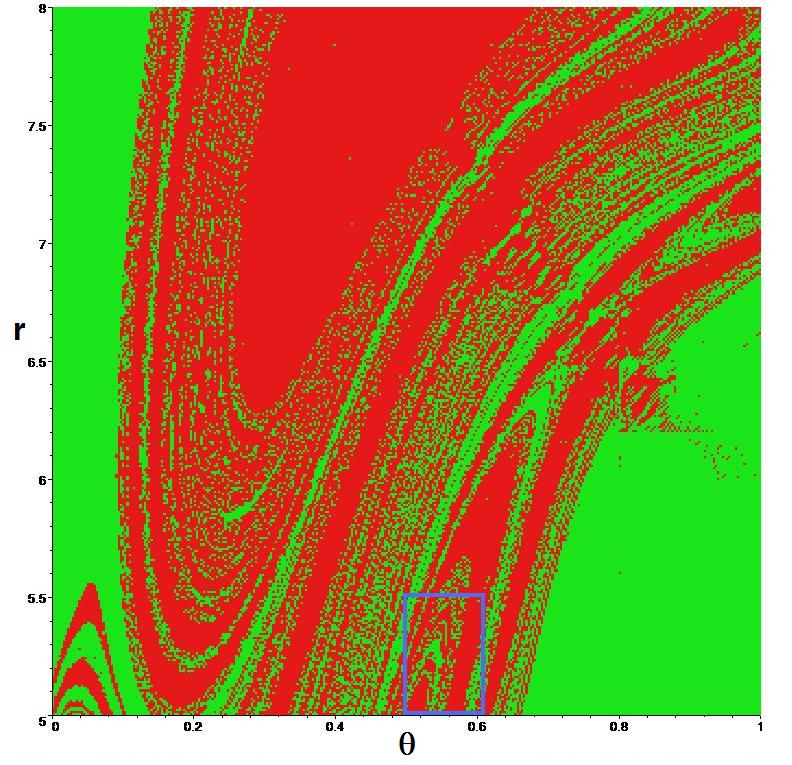}
\end{center}
\end{figure}

In figure \ref{fig:BA0}
 we plot the basins of attraction in a large subset of our phase space. All these initial conditions correspond to the values of $\theta$ and $r$ shown in the figure, $p_r=0$ and $p_\theta$ given by the constrain $H_r=0$.
Each figure contains $325 \times 325$ points.
Red points correspond to the outcome of the geodesic crossing into the event horizon, blue points to escaping to infinity, while green points correspond to the oscillatory chaotic behavior as in figure \ref{fig:rvst}.
The condition for capture is simply $r\leq r_H$.
The conditions for escape were defined as in section \ref{sssec:fp} and correspond to  $r_*\geq 100r_H$.
For green points we consider trajectories that,  after $5000$ iterations, neither were captured nor escaped to infinity.

\begin{figure}
\caption{Basins of attraction. A zoom of the region bounded by the blue rectangle in fig.\ref{fig:BA0}.}
\label{fig:BA1}
\begin{center}
\includegraphics[width=.85\textwidth]{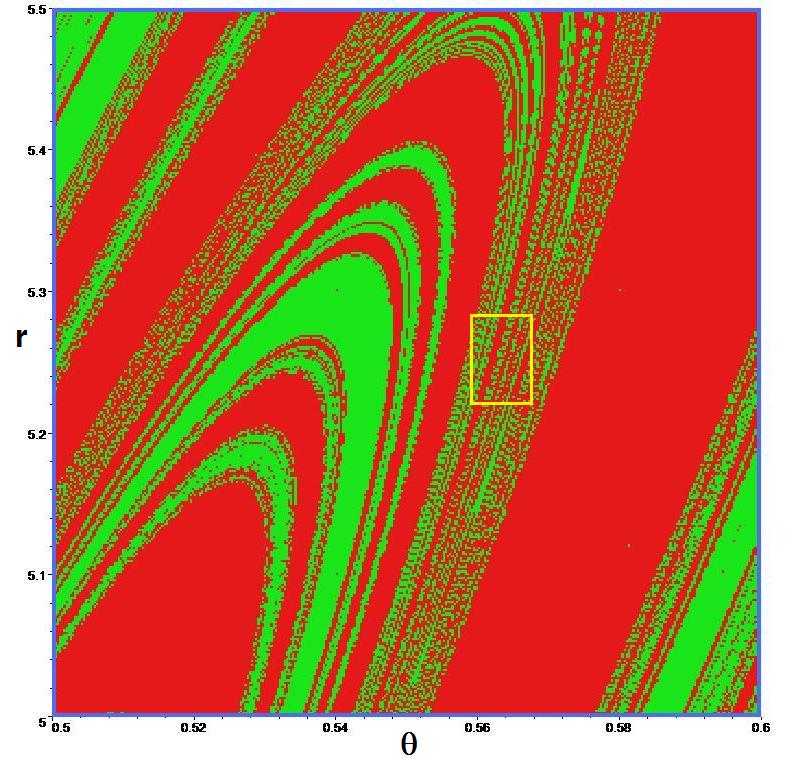}
\end{center}
\end{figure}

\begin{figure}
\caption{Basins of attraction. A zoom of the region bounded by the yellow rectangle in fig.\ref{fig:BA1}.}
\label{fig:BA2}
\begin{center}
\includegraphics[width=.85\textwidth]{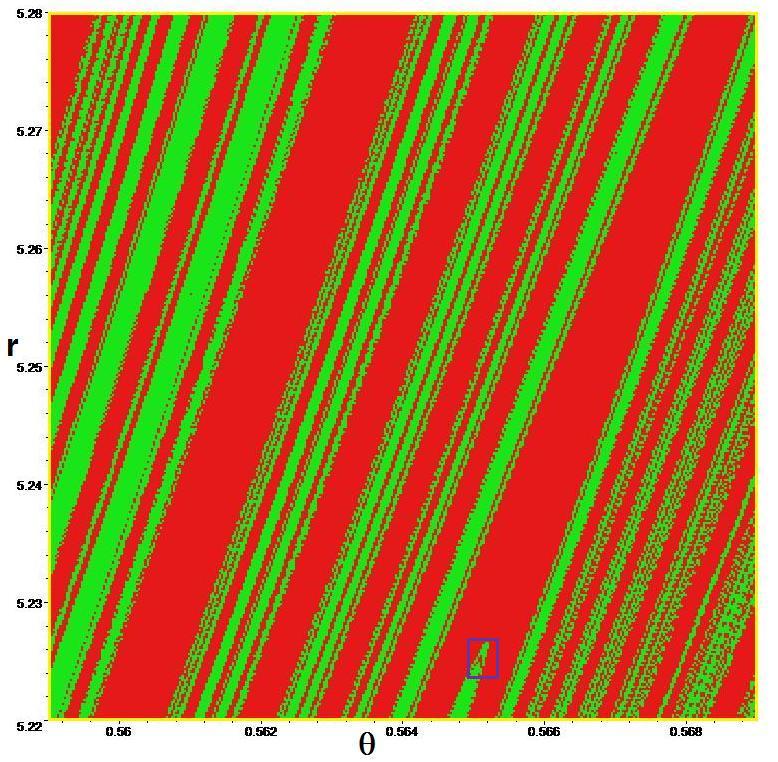}
\end{center}
\end{figure}

There are some well defined red and green regions in figure \ref{fig:BA0}. As we already noted, in our case, initial conditions yielding escape to infinity form a zero measure set.
The red and green zones are divided by boundaries which look fuzzy, as if containing finer structure.
To verify this, in figure \ref{fig:BA1}
we did a zoom of the region bounded by the blue rectangle in figure \ref{fig:BA0}.
We have found that, indeed, our description of figure \ref{fig:BA0} is also valid for the zoomed region.
Furthermore, we did a zoom of the region bounded by the yellow rectangle in figure \ref{fig:BA1}, which is presented in figure \ref{fig:BA2}.
We see that finer and finer structure arises as we zoom into the boundaries.
Moreover if, for instance, we zoom now the region bounded by the purple rectangle in figure \ref{fig:BA2} we will obtain something similar to figure \ref{fig:BA1}. This way we have verified the self-similar fine structure of the basins boundaries, a clear evidence for their fractality.
Crucially for us, there are areas where an arbitrarily small change in initial conditions leads to a completely different outcome.

\section{Field theory interpretation}\label{field}

\subsection{Identifying the corresponding operator}

In this section we argue that the operators being described by the ring string configuration discussed in this paper are generalizations of the Gubser-Klebanov-Polyakov (GKP) operators described in \cite{Gubser:2002tv}. Let us, for a moment, assume that there is no black hole in our background, that is, that we are considering the ring string in pure $AdS_5$ which correspond to $M=0$ in the metric described by equation  (\ref{eq:metric}). We will further introduce a new radial coordinate $\rho$ of the form $r=b\sinh\rho$, where $b$ is the radius of $AdS_5$ as described in equation (\ref{eq:metric}). The metric is then the standard $AdS_5$ in global coordinates:
\be
ds^2 =-\cosh^2\rho dt^2 +d\rho^2 +\sinh^2\rho(d\theta^2 +\cos^2\theta d\phi^2 +\sin^2\theta d\psi^2).
\ee
We can consider the same embedding of the ring string in this background as described in equation (\ref{eq:embedding}). We end up with the following  Nambu-Goto Lagrangian
\be
{\cal L}= \alpha \, \sin\theta \sinh\rho \sqrt{(\dot{t})^2\cosh^2\rho -\dot{\rho}^2 -\sinh^2\rho (\dot{\theta}^2 + \cos^2\theta (\dot{\phi})^2)}.
\ee
The key conclusion we want to draw from this exercise is that the spacetime energy and angular momentum are
\bea
\Delta &\sim &\int d\sigma d\tau \frac{\dot{t}\,\, \sin\theta \sinh\rho\,\cosh^2\rho }{\sqrt{(\dot{t})^2\cosh^2\rho -\dot{\rho}^2 - (\dot{\theta}^2 + \cos^2\theta (\dot{\phi})^2)\sinh^2\rho}} \nonumber \\
S &\sim &\int d\sigma d\tau  \frac{ \dot{\phi}\,\,\sin\theta \cos^2\theta \,\sinh\rho \sinh^2\rho}{\sqrt{(\dot{t})^2\cosh^2\rho -\dot{\rho}^2 - (\dot{\theta}^2 + \cos^2\theta (\dot{\phi})^2)\sinh^2\rho}} \nonumber \\
\eea
These expressions are simple generalizations of those appearing in \cite{Gubser:2002tv}. We thus, hope that our operators are generalizations of GKP operators. One key observation is in order. The derivatives in our cases are somehow exchanged, namely our coordinate $\rho$ depends nontrivially on $\tau$ while the dependence of $\rho$ in GKP was nontrivial in $\sigma$.  We can speculate that the $\sigma \leftrightarrow \tau$ switch is a sort of worldsheet duality which we loosely associate with $S$ modular transformation on the string side.

It was already clear in the case of GKP \cite{Gubser:2002tv} that the precise operator dual to the string is hard to write down explicitly; in particular, its form depends on the parameters used: angular velocity or range of the coordinate $\rho$. More precisely, the string configuration in GKP interpolates between typical Regge states for small size of the spinning string and twist-two operators when the length of the string grows parametrically. We, therefore, will simply note that in some regime, the general form of the operators dual to the string configuration we are discussing should contain, as building blocks,  operators of the form
\be
{\cal O}_{\Delta, S}\sim {\rm Tr} \Phi^I (D^{\leftrightarrow})^S\Phi^I.
\ee
More generally we can consider some insertion into this operator of the form
\bea
{\cal O}_{k}^{\Delta, S}& \sim  &{\rm Tr} \Phi^I D_{(\mu_1}\ldots D_{\mu_k}\Phi^m \ldots D_{\mu_S)}\Phi^I.
\eea
The above operator, is not a singlet with respect to $SO(6)$ but we are only concerned with the approximate general structure. One can even, attempt to implement momentum conservation along the string as done in \cite{Berenstein:2002jq}. At the expense of being overly speculative we conjecture that, in some regime of parameters, the ring string of our analysis is dual to an operator of the general form
\bea
{\cal O}&\sim & \sum\limits_{n}\sum\limits_{k=1}^{S}e^{\frac{2\pi i n k}{S}}{\cal O}_{k}^{\Delta, S}.
\eea
In the context of the ring string in the Schwarzschild black hole background the situation is even more complicated. It is very likely that the state we are describing contains an averaging in the canonical ensemble over a large number of operators. We will not pursue this issue further in this paper.

\subsection{Largest Lyapunov Exponents and Poincar\'e Recurrences}

A key aspect in the analysis of the largest Lyapunov exponent is the explicit assumption that one tracks the distance between two nearby trajectories at late times. Algorithmically, it means that the early time part of the trajectories is discarded form the computation.

We find an interesting analogy with the fact that it has been long suggested that many aspects of black hole physics require an understanding of late times evolution. The gauge/gravity correspondence naturally extends  to the context of black hole physics. In principle, given this equivalence of a gravity theory (containing black holes) and a field theory, unitarity of field theories provides a possible resolution for black hole puzzles like information loss. Maldacena suggested to follow the detailed very long time properties of correlation functions in \cite{Maldacena:2001kr}.

Let us recall the structure of correlation functions in generic field theories. Initial perturbation (insertion of operators) of a given thermal system will be damped by thermal dissipation as long as the time scale is too short to resolve possible gaps in the spectrum (Heisenberg times). For a unitary system with a discrete spectrum the perturbation is prevented from dying out. In fact, for any required precision there will exist a time for which the correlation functions returns to its initial value within the required precision  \cite{Thirring}.

Let us be a bit more precise, following the presentation, for example, of \cite{Barbon:2003aq}. As mentioned before, Poincare recurrences are characteristic of finite bounded systems that evolve in a unitary way. For  bounded system with discrete energy one considers an arbitrary operator $A$ which evolves in time as $A(t)=e^{itH}A(0)e^{-itH}$. The corresponding matrix elements can be computed in the energy basis as
\be
G_E(t)=e^{-S(E)}\sum\limits_{E_i,E_j\le E}|A_{ij}|^2e^{i(E_i-E_j)t}.
\ee
The object of study is the function $G_E(t)$ which is a correlation function. The key observation is that, if we define Heisenberg time as $t_H\equiv 1/\omega$, where $\omega ={\rm Inf} (E_i-E_j)$, then this is the time scale that reflects the discreteness of the spectrum. For $t\ll t_H$ the spectrum is approximately continuous.

If the matrix elements of some operator $A$ in the energy basis have frequency with $\Gamma$, the correlator will decay with characteristic lifetime of order $\Gamma^{-1}$: $G_E(t)$ Standard dissipative behavior in $\Gamma^{-1}\ll t\ll t_H$. For $t>t_H$ most phases in $G_E(t)$ would have completed a period and the function $G_E(t)$ starts  showing irregularities; it is a quasi-periodic function of time. Despite thermal damping, it returns arbitrarily close to the initial value over periods of the order of the recurrence time.

In a very clear and pedagogical work Barb\'on and Rabinovici \cite{Barbon:2003aq} clarified the importance of the distinction between discrete and continuous spectrum.  In particular, they highlighted the fact that the presence of a horizon implies a continuous spectrum and thus, in that limit, Poincar\'e recurrences should not be observed.

What is the field theory object or operation dual to the computation of the Lyapunov exponent? We argue that the distance between two string worlsheets is related to the correlation function of the corresponding operators that the worldsheets are dual to. This intuition arises from the a similar situation with Wilson loops discussed in   \cite{Berenstein:1998ij}. More generally, the correlator between two worldsheets is, in principle, a third worldsheet. In our case we believe that such worldsheet is approximated by the distance between the two original woldsheets.

Now, by definition, the Lyapunov exponent describes the late time behavior of the distance between two trajectories. We therefore would like to translate this result as the late time behavior of a correlation between the two operators described by the worldsheets in question. Moreover, assuming that the late-time behavior of the correlators is dictated by the Poincar\'e recurrences we conclude with the suggestion that the Poincar\'e
recurrence time $t_{PR}$ be related to the largest Lyapunov exponent  $\lambda$ as
\bea
t_{PR}=\frac{1}{\lambda}.
\eea
One last caveat is in order. The largest Lyapunov exponent is computed in phase space, therefore we essentially find a lower bound for the Poincar\'e recurrence time.

Late time correlation functions are the central object in the Information Loss paradox.  The argument above ties Poincare recurrences to unitarity in the presence of a discrete spectrum. Thus, the relationship between chaos and its description in terms of the Lyapunov exponent in string theory might point to an underlying restoration of unitarity.

As in the cases of \cite{Berenstein:2002jq} and \cite{Gubser:2002tv}, we believe the reason we are able to observed this, otherwise very small effect is due to the enhancing effect of large quantum numbers, in this case the spin $S$. Schematically we have: $e^{-N^2}\longrightarrow e^{-N^2/S^2 }\sim {\cal O}(1)$, that is, what would have been a suppressed effect gets enhanced thanks to the presence of a large quantum number.

It is worth mentioning the academic character of the above speculation. The problem with recurrence times is that {\it typically} they are so unimaginably long that they have no physical relevance.  Think of the factors $e^{i\omega_j t}$ as clocks, and let there be $N$ of them. The question of recurrence is now the question of how long it takes for a certain configuration of clocks to reappear to within some angular accuracy $\Delta \phi$. The recurrence time is $(\Delta \phi/2\pi)^{-N}/\omega$, where $1/\omega$ is the average of $1/\omega_j$.  Note that for just $N=10$, $1/\omega=1$sec., and $\Delta/2\pi= 1/100$ (an accuracy of 1\%), the recurrence time is $10^{20}$sec. which is much longer than the age of the universe \cite{Thirring}.

\section{Conclusions}
In this paper we have presented a detailed analysis of the phase space structure of a ring string in the Schwarzschild black hole in $AdS_5$ space. We have found convincing evidence in support of chaotic behavior. As is usually the case in bounded nonintegrable dynamical systems, we expect this dynamics to be generic.

The general question that we are starting to address is: {\it What is the interpretation, on the dual field theory side, of the main quantifiers of chaotic behavior?} In other words, we try to extend the AdS/CFT dictionary to include entries related to chaotic dynamics. We have proposed that one of the key indicators  of chaos - a positive largest Lyapunov exponent -- be identified, on the dual field theoretic side, with the appropriate bound for the time scale of Poincar\'e recurrences. Many other indicators of chaos remained to be discussed, for example, the interpretation  of various fractal dimensions which are very characteristic of chaotic systems and for which we found evidence in the study of basins of attraction. Similarly, the interpretation of some of the concrete indicators we computed in this paper have not been addressed: the power spectrum, the Poincar\'e sections and the basins of attraction.

We finish by stating a number of open questions that seem particularly interesting to us, these questions are of a more technical nature and do not necessarily depend on the conceptual framework given by the AdS/CFT correspondence. The standard approach to dynamical systems naturally specializes to the case of point particles. To our modest knowledge, not much is known or written about strings. There are substantial differences, most notably, the Virasoro constraints. It would be interesting to study and clarify the role of strings, or more generally extended objects (p-branes) using the arsenal of techniques available for dynamical systems.

Integrability, understood as the ability to write the equations of motion in quadrature, is usually assumed to imply the absence of chaos in dynamical systems. This intuition, again, is rooted in the particle interpretation. In the case of motion of strings we are now talking about integrability of systems with infinitely many degrees of freedom. The string has infinitely many degrees of freedom as it can oscillate in infinitely many modes $\exp(i n \sigma)$ and integrability of the strings gives infinitely many conserved quantities. Is this as restrictive as in the case of a finite number of degrees of freedom where one can use the action/angle variables to completely characterize the system? There is a precedent to this question that might indicate that the answer is rather elaborate. Namely, the Toda system is integrable, however, it has been shown that any polynomial truncation of the Toda potential displays chaotic behavior. A behavior  along these lines might occur in the case of the string. If we allow it to have all the excitations $\exp(i n \sigma)$ turned on, it could behave very differently from situations in which we freeze some of the excitations by virtue of  Ans\"atze similar to the one considered here.

Another very interesting aspect is the role of the asymptotic conditions, that is, asymptotically flat versus asymptotically $AdS$ \cite{WholeStudy}. For example, in a previous study of a ring string in asymptotically Minkowski space time \cite{Frolov:1999pj}, the structure of the basins of attraction seem to be different from the one obtained in our case. More generally, we would like to cleanly pinpoint the role of asymptotically $AdS$ spaces.

\section*{Acknowledgments}
We thank, J. Barb\'on, B. Burrington, A. Frolov, D. Minic, the participants of the XII Mexican for comments, in particular J. Antonio Garc\'{\i}a.
We are very especially grateful to J. Barb\'on, B. Burrington and M. Kruczenski for reading the manuscript and providing feedback. We are particularly thankful to Tim Raben, who has been collaborating with us on very related topics.
L. P-Z. thanks the KITP in Santa Barbara for hospitally during the latest stages  of this work. This research was supported in part by the National Science Foundation under Grant NSF PHY05-51164. This work is  partially supported by Department of Energy under grant DE-FG02-95ER40899 to the University of Michigan.
C. T-E. thanks the MCTP in Ann Arbor for hospitally during the latest stages  of this work.


\begin{thebibliography}{99}


\bibitem{Maldacena:1997re}
  J.~M.~Maldacena,
  ``The large N limit of superconformal field theories and supergravity,''
  Adv.\ Theor.\ Math.\ Phys.\  {\bf 2} (1998) 231
  [Int.\ J.\ Theor.\ Phys.\  {\bf 38} (1999) 1113]
  [arXiv:hep-th/9711200].

\bibitem{Witten:1998qj}
  E.~Witten,
  ``Anti-de Sitter space and holography,''
  Adv.\ Theor.\ Math.\ Phys.\  {\bf 2} (1998) 253
  [arXiv:hep-th/9802150].

\bibitem{Gubser:1998bc}
  S.~S.~Gubser, I.~R.~Klebanov and A.~M.~Polyakov,
  ``Gauge theory correlators from non-critical string theory,''
  Phys.\ Lett.\  B {\bf 428} (1998) 105
  [arXiv:hep-th/9802109].

\bibitem{Aharony:1999ti}
  O.~Aharony, S.~S.~Gubser, J.~M.~Maldacena, H.~Ooguri and Y.~Oz,
 ``Large N field theories, string theory and gravity,''
  Phys.\ Rept.\  {\bf 323}, 183 (2000)
  [arXiv:hep-th/9905111].

\bibitem{Berenstein:2002jq}
  D.~E.~Berenstein, J.~M.~Maldacena and H.~S.~Nastase,
  ``Strings in flat space and pp waves from N = 4 super Yang Mills,''
  JHEP {\bf 0204}, 013 (2002)
  [arXiv:hep-th/0202021].

\bibitem{Gubser:2002tv}
  S.~S.~Gubser, I.~R.~Klebanov and A.~M.~Polyakov,
  ``A semi-classical limit of the gauge/string correspondence,''
  Nucl.\ Phys.\  B {\bf 636} (2002) 99
  [arXiv:hep-th/0204051].


\bibitem{Hadamard}
J. Hadamard,``Les surfaces \`a courbure oppos\'ees et leurs lignes g\'eodesiques,''  J. Math. Pures et Appl. 4 (1898) 27.

\bibitem{anosov}
D.~V.~Anosov, ``Geodesic Flows on Closed Riemann Manifolds with Negative Curvature,'' Proc. Steklov Institute of Mathematics, vol. {\bf 90}, 1967.

\bibitem{Arnold}
V.~I.~Arnol'd, ``Mathematical Methods of Classical Mechanics,'' Springer-Verlag, 1989.

\bibitem{Hawking:1973uf}
  S.~W.~Hawking and G.~F.~R.~Ellis,
{\it  Cambridge University Press, Cambridge, 1973}

\bibitem{Frolov:1999pj}
  A.~V.~Frolov and A.~L.~Larsen,
  ``Chaotic scattering and capture of strings by black hole,''
  Class.\ Quant.\ Grav.\  {\bf 16} (1999) 3717
  [arXiv:gr-qc/9908039].

\bibitem{Enns}
R.~H.~Enns and G.~C.~McGuire, ``Nonlinear Physics with MAPLE for Scientists and Engineers. Second Edition,'' Meas. Sci. Technol. 12 2022, 2001.

\bibitem{jcsprott}
J. C. Sprott, ``Chaos and Time-Series Analysis,'' Oxford University Press, 2003.

\bibitem{Ott}
E. Ott, ``Chaos in Dynamical Systems,'' Cambridge University Press, Second Edition 2002

\bibitem{Brown:1996}
  R.~Brown and L.~O.~Chua,
  ``Clarifying chaos: examples and counterexamples,''
Int.\ J.\ Bifurcation and Chaos {\bf 6} (1996) 219

\bibitem{Brown:1998}
  R.~Brown and L.~O.~Chua,
  ``Clarifying Chaos II: Bernoulli Chaos, Zero Lyapunov Exponents and Strange Attractors,''
Int.\ J.\ Bifurcation and Chaos {\bf 8} (1998) 1

\bibitem{kolmogorov}
A.N. Kolmogorov, ``A New Metric Invariant of Transitive Dynamical Systems and Automorphisms in Lebesgue Spaces,'' Dokl. Acad. Nauk SSSR {\bf 119}, 861 (1958).

\bibitem{Osedelec}
  V.~I.~{Osedelec},
  ``A Multiplicative Ergodic Theorem: Lyapunov Characteristic Numbers for Dynamical Systems,''
 Trans.\ Moscow Math.\ Soc.\ {\bf 19} (1968) 197


\bibitem{Motter:2003jm}
  A.~E.~Motter,
  ``Relativistic chaos is coordinate invariant,''
  Phys.\ Rev.\ Lett.\  {\bf 91} (2003) 231101
  [arXiv:gr-qc/0305020].

\bibitem{Motter:2009kv}
  A.~E.~Motter and A.~Saa,
  Phys.\ Rev.\ Lett.\  {\bf 102}, 184101 (2009)
  [arXiv:0903.2296 [nlin.CD]].

\bibitem{sprott1}
J. C. Sprott, ``Numerical Calculation of Largest Lyapunov Exponent,'' http://sprott.physics.wisc.edu/chaos/lyapexp.htm


\bibitem{McDonald:1985}
  S.~W.~McDonald, C.~Grebogi, E.~Ott and J.~A.~Yorke,
  ``Fractal Basin Boundaries,''
  Physica D {\bf 7}, 125  (1985)


\bibitem{Cornish:1997iw}
  N.~J.~Cornish,
  ``Fractals and symbolic dynamics as invariant descriptors of chaos in
  general relativity,''
  arXiv:gr-qc/9709036.


\bibitem{Maldacena:2001kr}
  J.~M.~Maldacena,
  ``Eternal black holes in Anti-de-Sitter,''
  JHEP {\bf 0304} (2003) 021
  [arXiv:hep-th/0106112].

\bibitem{Thirring}
W. Thirring, ``Quantum Mathematical Physics: Atoms, Molecules and Large Systems,'' Springer 2002.


\bibitem{Barbon:2003aq}
  J.~L.~F.~Barbon and E.~Rabinovici,
  ``Very long time scales and black hole thermal equilibrium,''
  JHEP {\bf 0311} (2003) 047
  [arXiv:hep-th/0308063].

\bibitem{Berenstein:1998ij}
  D.~E.~Berenstein, R.~Corrado, W.~Fischler and J.~M.~Maldacena,
  ``The operator product expansion for Wilson loops and surfaces in the  large
  N limit,''
  Phys.\ Rev.\  D {\bf 59} (1999) 105023
  [arXiv:hep-th/9809188].


\bibitem{WholeStudy}
L. A. Pando Zayas, T. Raben and C. A. Terrero-Escalante, {\it In preparation.}








\end{thebibliography}
\end{document}